\documentclass[showpacs,preprintnumbers,amsmath,nofootinbib,aps]{revtex4}

\usepackage{graphicx,epsf, epsfig, amssymb, multirow}% Include figure files
\usepackage{bm}
\usepackage{amsmath}

\def\missPT{${P\mkern-11mu\slash}_T$}
\def\bentarrow{\:\raisebox{1.3ex}{\rlap{$\vert$}}\!\rightarrow}
                                                                                                                           
\begin{document}
                                                                                                                           
\title{Discriminating Supersymmetry and Black Holes at the Large Hadron Collider}

\author{Arunava Roy}
\email{arunav@olemiss.edu}
\affiliation{Department of Physics and Astronomy, University of Mississippi,
University, MS 38677-1848, USA}
                                                                                                                           
\author{Marco Cavagli\`a}
\email{cavaglia@olemiss.edu}
\affiliation{Department of Physics and Astronomy, University of Mississippi,
University, MS 38677-1848, USA}

\date{\today}

\begin{abstract}
We show how to differentiate the minimal supersymmetric extension of the standard model from black
hole events at the Large Hadron Collider. Black holes are simulated with the \texttt{CATFISH}
generator. Supersymmetry simulations use a combination of \texttt{PYTHIA} and \texttt{ISAJET}. Our
study, based on event shape variables, visible and missing momenta, and analysis of dilepton
events, demonstrates that supersymmetry and black hole events at the LHC can be easily
discriminated. 
\end{abstract}
\maketitle

\section{Introduction\label{intro}}

At CERN's Large Hadron Collider (LHC) \cite{CERN web} protons will soon collide at an astonishing
800 million times per second to provide experimental evidence for the Higgs
\cite{Higgs:1964ia,ATLAS:1999}, supersymmetry (SUSY) \cite{Wess:1973kz,Baer:1995nq} or extra
dimensions \cite{Arkani-Hamed:1998rs,Randall:1999ee,Appelquist:2000nn}. SUSY is widely considered
to be one of the best candidates for physics beyond  the Standard Model (SM). It provides an
explanation for the Higgs mass problem, a candidate for cold dark matter, and unification of low
energy gauge couplings by introducing superpartners to SM fields (see Ref.~\cite{Martin:1997ns}
and  references therein).  An alternative to SUSY is given by phenomenological extra-dimensional
models such as Large Extra Dimensions (LEDs) \cite{Arkani-Hamed:1998rs}, warped braneworlds
\cite{Randall:1999ee} or universal extra dimensions \cite{Appelquist:2000nn}. Scenarios with LEDs
are specially appealing. In these models, gravity becomes strong at the TeV scale, where radiative
stability is achieved. The fundamental scale of gravity, $M_\star\sim 1$ TeV, is related to the
observed Planck scale, $M_{\rm Pl}$, by the relation $M_{\rm Pl}^2\sim V_n M_{\star}^{n+2}$, where
$V_n$ is the volume of the extra $n$-dimensional space. One of the most astounding consequences of
the existence of extra-dimensions would be the  production of subatomic Black Holes (BHs) in
particle colliders \cite{Argyres:1998qn,Ahn:2002zn} and cosmic ray showers \cite{Feng:2001ib}. (For
reviews, see Refs.~\cite{Cavaglia:2002si}.) 

The ATLAS \cite{Armstrong1:1994} and CMS \cite{Armstrong2:1994} experiments at the LHC are
entrusted with the task of studying events with large transverse momentum ($P_T$), a signature
common to both SUSY and extra dimensions. While we wait for these experiments to start collecting
data, it is worthwhile to look into means of distinguishing SUSY and extra-dimensional models
\cite{Buescher:2006jm}. Comparisons of SUSY and universal extra dimensions/little Higgs models in
colliders have been investigated by various authors \cite{Rizzo:2001sd}. Discrimination of SUSY and
BH events by means of dilepton events was recently discussed by the authors in Ref.\
\cite{Roy:2007fx}. In this paper, we revisit that analysis and extend it to include event shape
variables, missing transverse momentum \missPT\ and visible energy. BH and SUSY events are
simulated with the BH generator \texttt{CATFISH} \cite{Cavaglia:2006uk} and the high-energy event
generator \texttt{PYTHIA} \cite{Sjostrand:2006za}, respectively. SUSY masses are set with
\texttt{ISAJET} \cite{Paige:2003mg}. The analysis below will show that SUSY and BH events can be
clearly distinguished at the LHC. BH events tend to be more spherical than SUSY events because of
the isotropic nature of BH decay. Thus event shape variables, such as sphericity, provide good
discriminators. On the contrary, visible energy and \missPT\ are less effective discriminators
because of the presence of invisible channels in both SUSY and BH models, which make the amount of
\missPT\ comparable in the two scenarios. The dilepton invariant mass is also an excellent
discriminator; the SUSY invariant mass shows a sharp cutoff at $\sim$ 100 GeV, which is absent in
the BH model because most of the dileptons originate from uncorrelated events.

The remainder of this paper is organized as follows. In Sect.~\ref{SUSYLHC} and \ref{BHLHC} we
briefly review the fundamentals of SUSY and TeV BHs which are needed for our analysis,
respectively. Simulations are described in Sect.~\ref{evt_sim}. The analysis of visible/missing
momentum and event shape variables is presented in Sect.~\ref{shape}, and the discrimination of
SUSY and BH using dileptons is discussed in Sect.~\ref{dilep}. Conclusions are presented in
Sect.~\ref{concl}.  
\section{Supersymmetry at the LHC\label{SUSYLHC}}
The Minimal Supersymmetric extension of the Standard Model (MSSM) \cite{Wess:1973kz} is the
simplest SUSY model. According to the MSSM, all SM fermions (bosons) must have a bosonic
(fermionic)  partner. Superpartners have identical masses, charges and quantum numbers of their SM
counterparts, differing only in their spin. The MSSM allows for the unification of electromagnetic,
weak and strong forces at $M_{GUT}\sim 10^{16}$ GeV. Since we do not observe superpartners of SM
particles at low energies, SUSY must be a broken symmetry. The SUSY breaking scale, i.e. the mass
scale at which we expect the first SUSY particles to appear, is generally assumed to be around 1
TeV. A method of SUSY breaking which is mediated by gravitational interactions is supergravity
(SUGRA). In its minimal version, mSUGRA is determined by a point in the five-dimensional moduli
space with parameters: 
\begin{itemize}
\item $m_0$, the common scalar mass at $M_{GUT}$;
\item $m_{1/2}$, the common gaugino mass at $M_{GUT}$;
\item $A_0$, the common trilinear coupling at $M_{GUT}$;
\item $\tan~\beta$, the ratio of the vacuum expectation values of the two Higgs fields; 
\item $\mu$, the sign of the Higgsino mass parameter. 
\end{itemize}
mSUGRA parameters for five typical LHC points are given in Table \ref{table1} \cite{Bartl:1996dr}.
Neutralino ($\tilde{\chi}_{i}^{0}$), gluino ($\tilde{g}$) and squark ($\tilde{q}$) masses are
determined by $m_0$ and $m_{1/2}$ as $\tilde{\chi}_{1}^{0}\sim m_{1/2}/2$, $\tilde{\chi}_{2}^{0}
\sim \tilde{\chi}_{1}^{\pm} \sim m_{1/2}$, $\tilde{g} \sim 3m_{1/2}$ and $m(\tilde{q}) \sim
({m_0}^2+6 m_{1/2}^2)^{1/2}$ \cite{Dittmar:1998rb}.  

\begin{table}[h]
\caption{Parameters for the five mSUGRA points discussed in the text.
The scalar mass and the gaugino mass are given in GeV.}
\begin{tabular*}{0.50\textwidth}%
{@{\extracolsep{\fill}}|c|cccccc|}
\hline
~LHC point~ & $m_0$ & $m_{1/2}$ & $A_0$ & tan $\beta$ & $\mu$ &\\
\hline
A & 100 & 300 & 300 & 2.1 & + &\\
\hline
B & 400 & 400 & 0 & 2 & +     &\\
\hline
C & 400 & 400 & 0 & 10 & +    &\\
\hline 
D & 200 & 100 & 0 & 2 & -     &\\
\hline
E & 800 & 200 & 0 & 10 & +    &\\
\hline
\end{tabular*}
\label{table1}
\end{table}

\begin{figure*}[ht]
\centerline{\null\hfill
    \includegraphics*[width=0.33\textwidth]{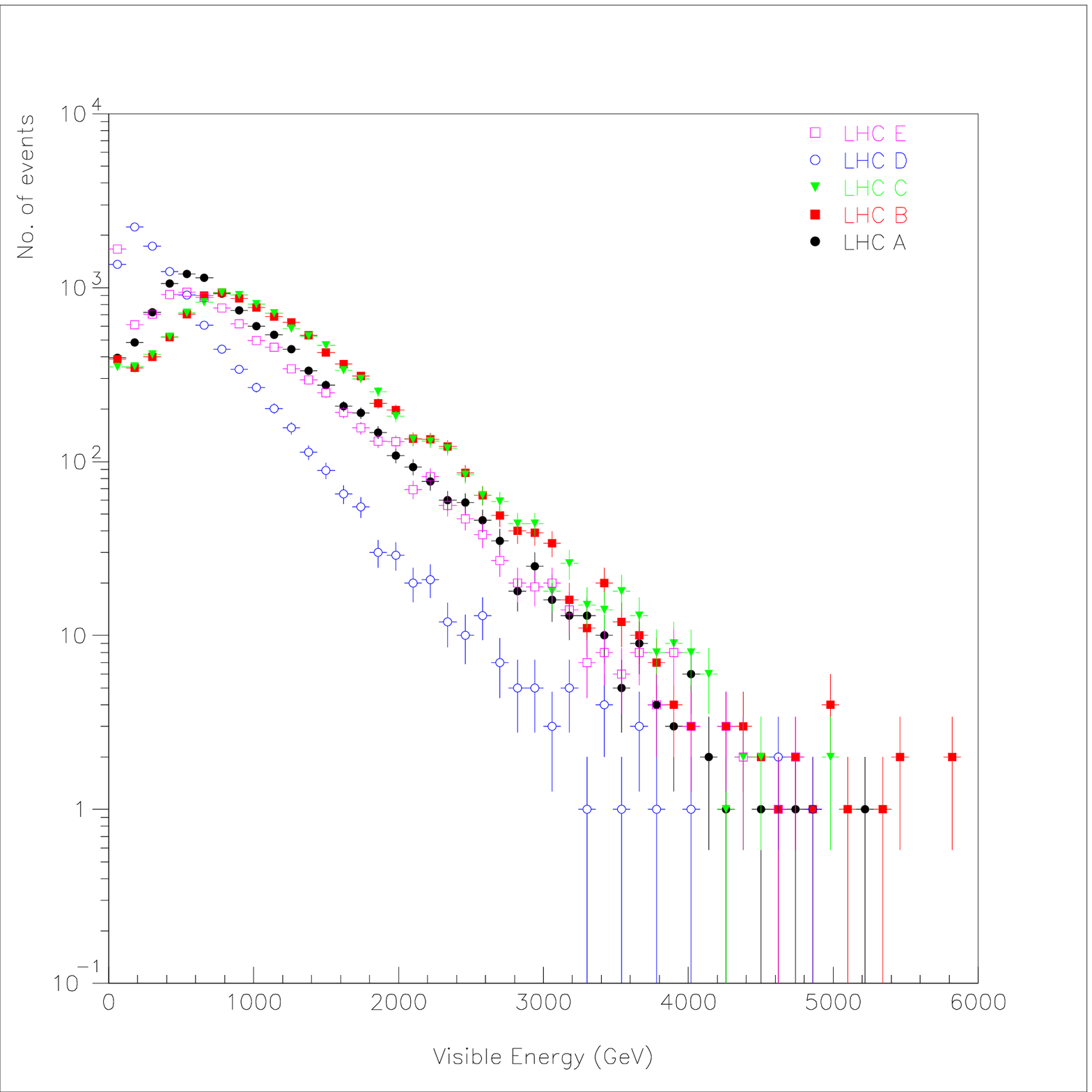}
    \hfill
    \includegraphics*[width=0.33\textwidth]{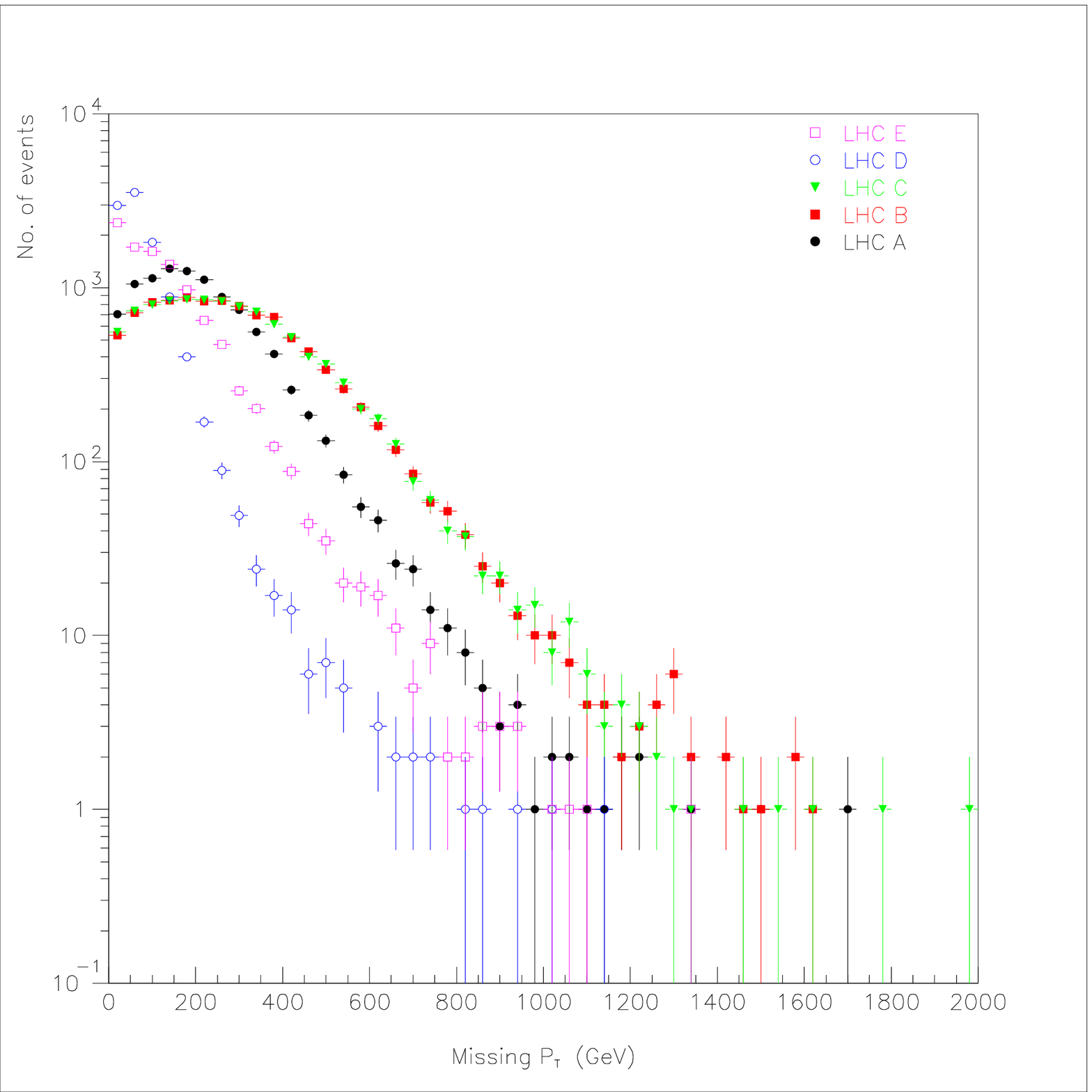}
    \hfill
    \includegraphics*[width=0.33\textwidth]{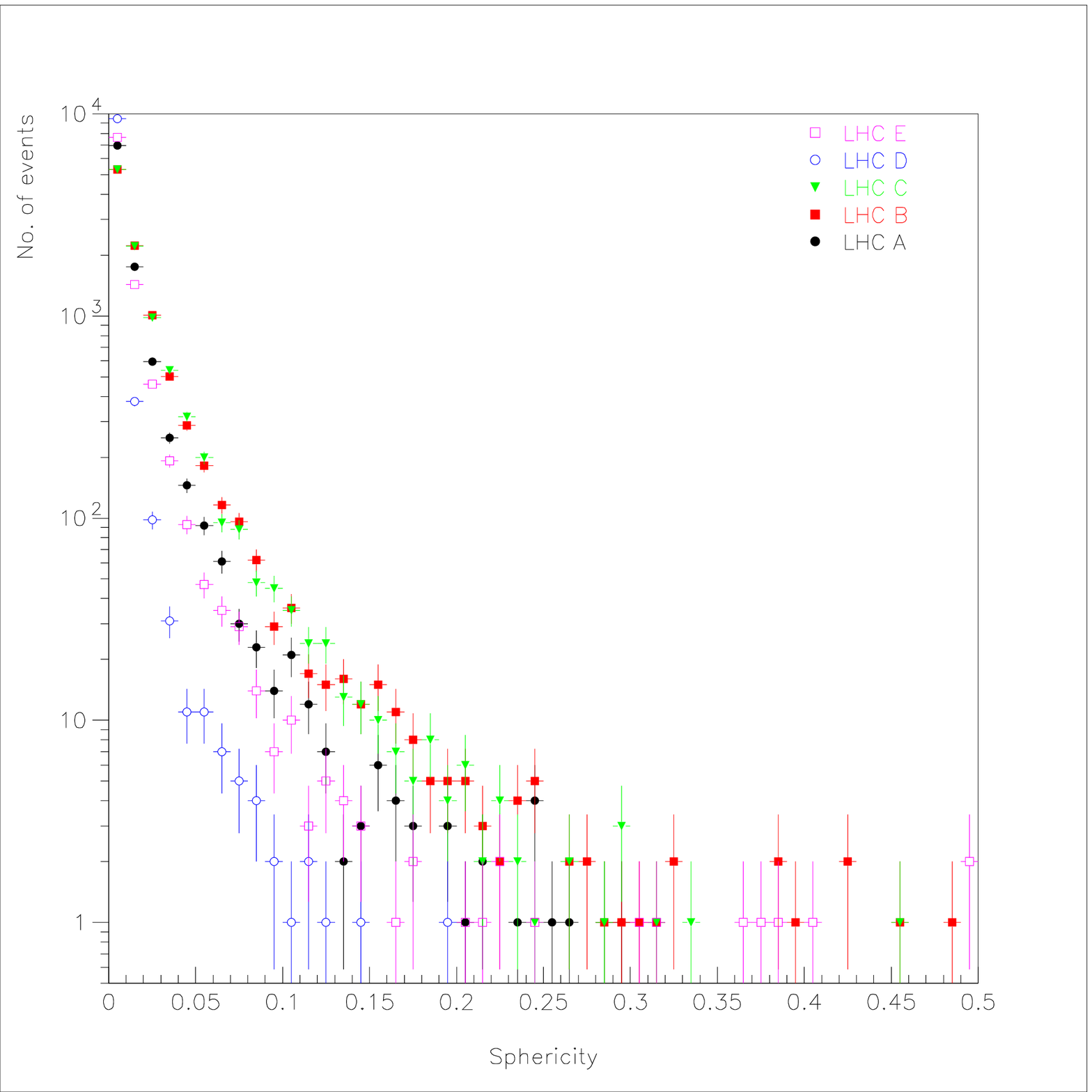}
    \null\hfill
    }
\caption{Comparison of visible energy (left), missing transverse momentum \missPT\ (middle) and
sphericity (right) for 10000 events for the five LHC points of Table \ref{table1} (A: black
filled circles, B: red filled squares, C: green filled triangles, D: blue open circles
and E: pink open squares).}
\label{fig1_1}
\end{figure*}

Visible energy, missing transverse momentum and sphericity for the five LHC points of Table
\ref{table1} are shown in Fig.~\ref{fig1_1}, where all SUSY processes except SM Higgs production
have been implemented. Sparticle production at point D (open blue circles) is higher as squarks and
gluinos are lighter. This point is usually taken as the comparison point between the LHC and other
experiments, e.g.\ Tevatron \cite{Arnowitt:1987hw} and NLC \cite{Feng:1995zd}. For the purposes of
our analysis, the difference between the five points is not significant and any of them can be
chosen as SUSY benchmark. In the following, we will consider point A. This is justified by the fact
that point A allows for SUSY Higgs production \cite{Hinchliffe:1996iu}. Since BHs may evaporate
into Higgs (see Sect.~\ref{BHLHC} below), a meaningful comparison of SUSY and BH events requires
the presence of the Higgs channel in both models. Moreover, distinguishability of SUSY and BH
events must be assessed by minimizing the differences between the two models. Since BH events are
characterized by up to several TeV of missing transverse momentum, SUSY points with large \missPT,
such as point A, must be considered.

A symmetry of the MSSM is $R$-parity \cite{Wess:1973kz}:
\begin{equation*}
P_{R}=(-1)^{3B+L+2s}\,,
\end{equation*}
where $B$ ($L$) is the baryon (lepton) number and $s$ is the particle spin. All SM particles have
$P_R=+1$ whereas their superpartners have $P_R=-1$. $R$-parity implies that SUSY particles are
always pair produced from SM particles. If $R$-parity is conserved, the endpoint of a SUSY process
at the LHC is a state with SM particles and two lightest stable SUSY particles (LSPs), which are
generally neutralinos. Being colorless and chargeless, the LSPs escape the detector and are the
source of missing transverse momentum, a leading signature of SUSY events. If $R$-parity is not
conserved, the missing transverse energy is reduced by the LSP decay. In the following, we will
assume that $R$-parity is conserved, in agreement with the MSSM (mSUGRA) scenario.

We end this section with a list of dominant SUSY interactions at LHC point A and the definition of
invariant mass. This is important for the following analysis because it enables us to select
processes that could serve as potential discriminators.
\begin{figure*}[ht]
\centerline{\null\hfill
    \includegraphics*[viewport=0 200 575 725,width=0.33\textwidth]{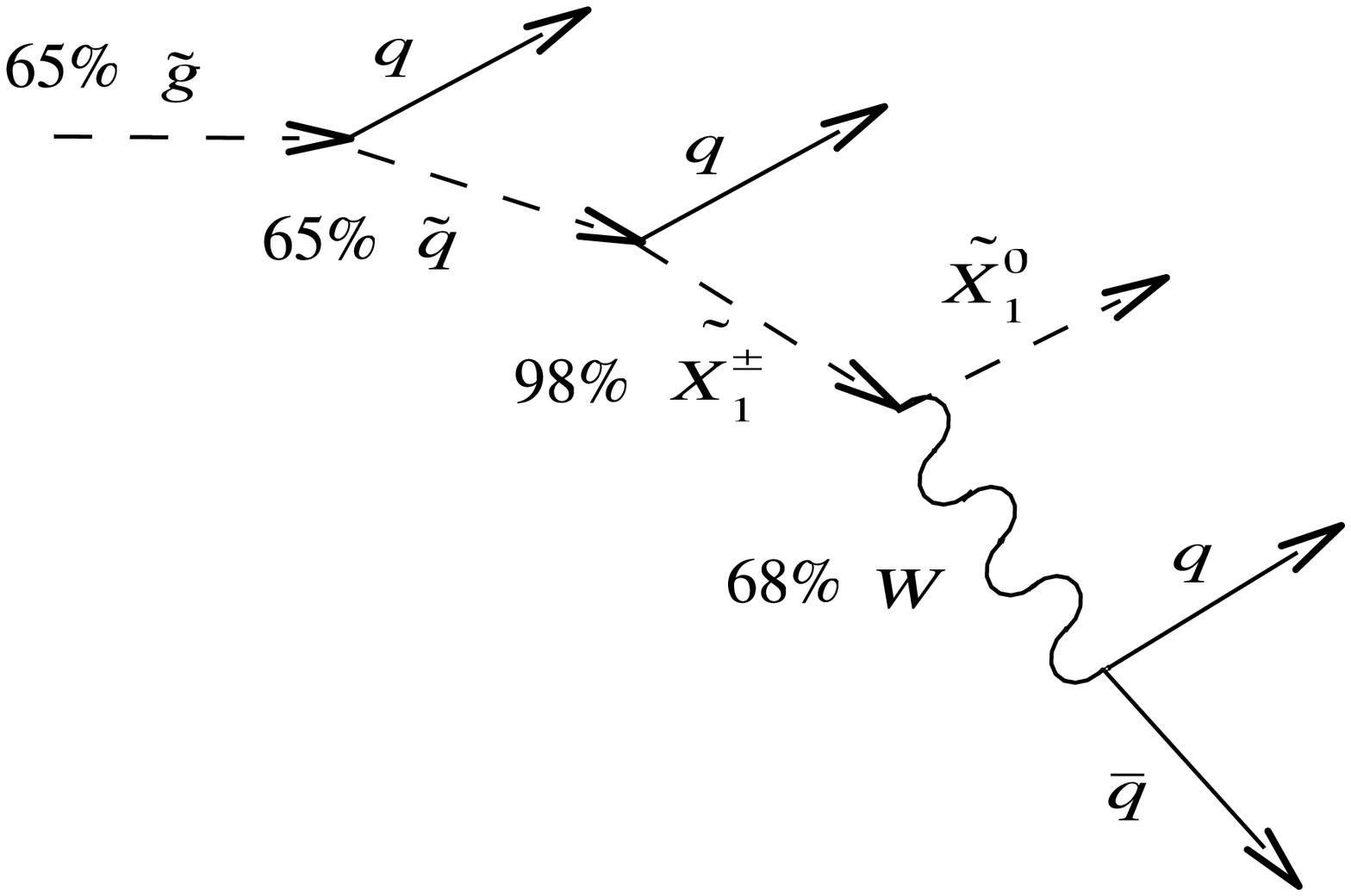}
    \hfill
    \includegraphics*[viewport=0 200 575 725,width=0.33\textwidth]{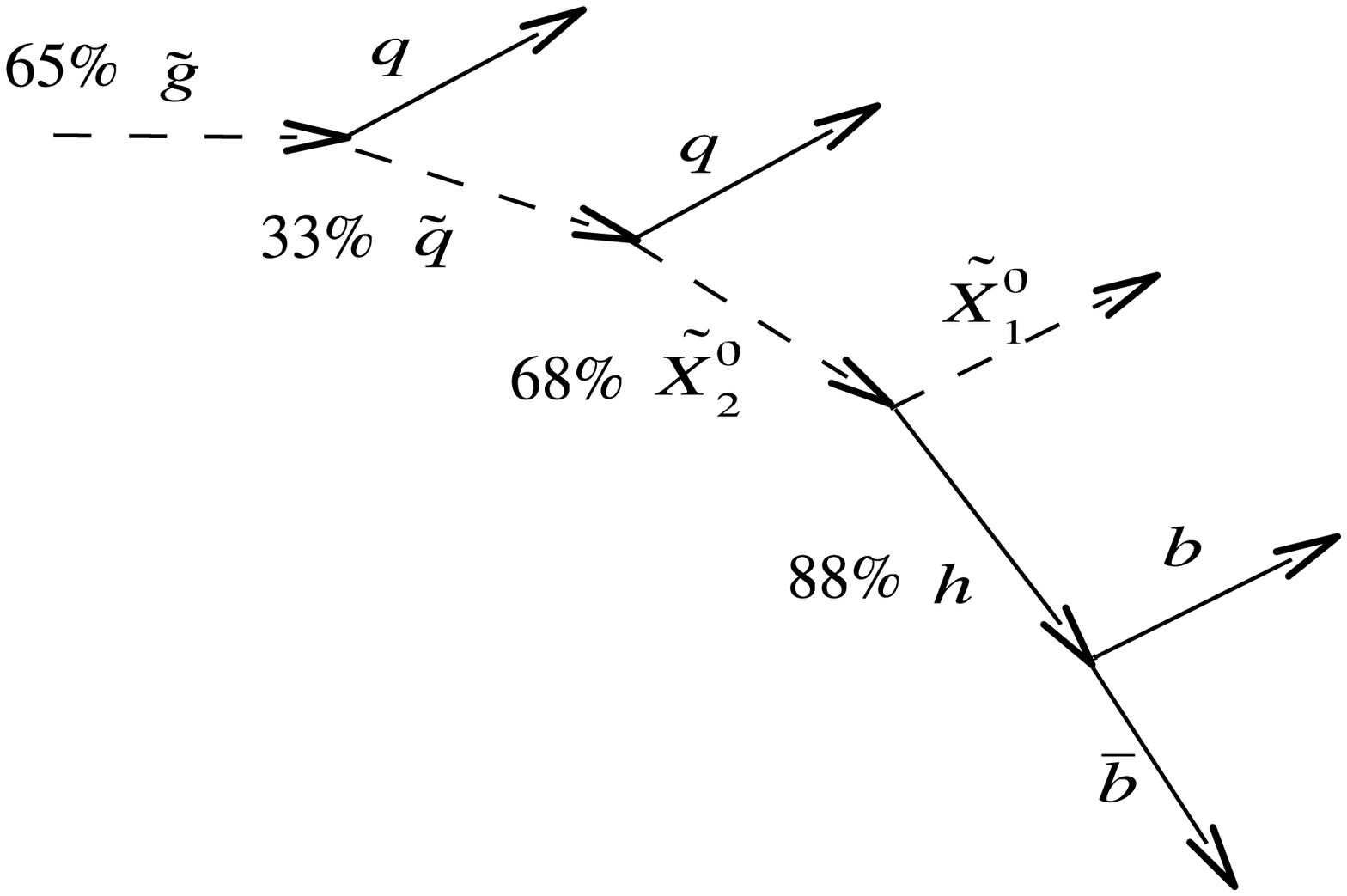}
    \hfill
    \includegraphics*[viewport=0 200 575 725,width=0.33\textwidth]{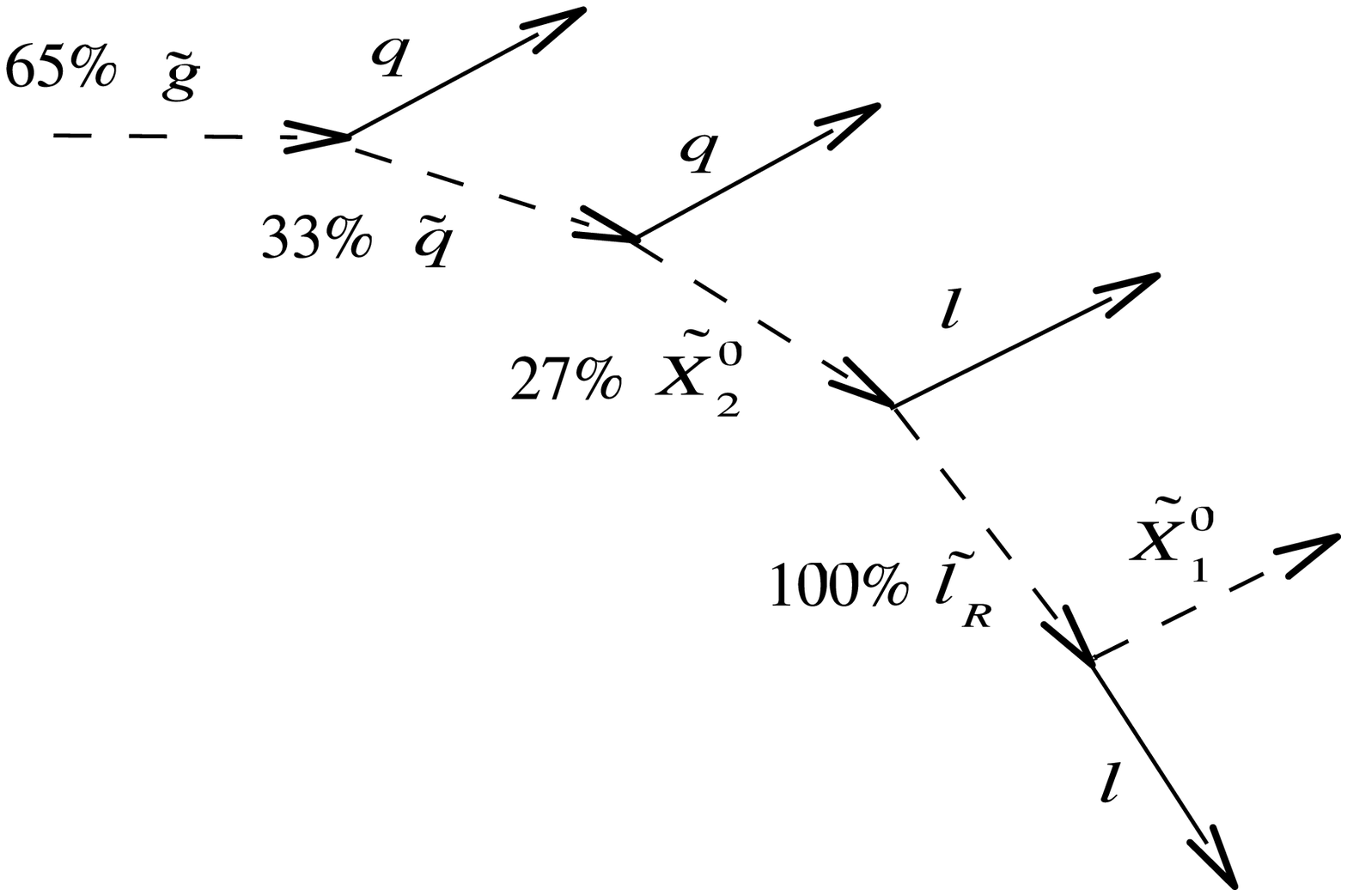}
    \null\hfill}
\caption{The top three SUSY decay chains and their branching ratios.} 
\label{DIAG}
\end{figure*}
The third decay chain in Fig.~\ref{DIAG} is specially interesting because it allows the separation
of isolated leptons from the hadronic background \cite{atlas111}. The invariant mass is defined as
\begin{equation*}
M_{12}=\sqrt{(E_1+E_2)^2-(\textbf{p}_1 +\textbf{p}_2)^2}=
\sqrt{2\textbf{p}_1\textbf{p}_2(1-\cos\theta)}\,,
\label{M12}
\end{equation*}
where $\theta$ is the angle between the two particles. The method of constructing invariant masses
from SUSY decay chains has been traditionally used to calculate sparticle masses
\cite{Hinchliffe:1996iu}. In Sect.\ \ref{dilep}  we will use the invariant mass of isolated
dileptons with large $P_T$ as a potential SUSY/BH discriminator. (For a review on lepton production
at colliders, see Ref.\ \cite{Diaconu:2004mj}.) In the rest frame of the second lightest
neutralino, the dilepton invariant mass is
\begin{equation*}
M_{ll}=\left[M_{\tilde{\chi}_2^0}^2+M_{\tilde{\chi}_1^0}^2-2~M_{\tilde{\chi}_2^0}~M_{\tilde{\chi}_1^0}
\left(1+\frac{P_{\tilde{\chi}_1^0}^2}{M_{\tilde{\chi}_1^0}^2}\right)\right]^{1/2}\,.
\label{invmass}
\end{equation*}
Since the momentum of the LSP is not constrained, the invariant mass distribution shows an edge at
$\sim 100$ GeV.
\section{Black holes at the LHC\label{BHLHC}}
In LED scenarios, $pp$ collisions at the LHC could produce TeV-mass BHs with characteristic
lifetimes of $10^{-25}$ seconds \cite{Cavaglia:2002si}. Numerous studies have focused on BH
signatures at the LHC \cite{Ahn:2002zn} and various Monte Carlo generators are available for
simulation purposes \cite{Cavaglia:2006uk,Dimopoulos:2001en,Harris:2003db} \footnote{Recently, a
new BH generator (BlackMax) appeared in the literature \cite{Dai:2007ki}. BlackMax is supposed to
include BH rotational effects. However, the gravitational loss for rotating BHs is artificially set
to zero. Since the energy loss due to gravitons is enhanced by rotation and extra-dimensional
effects, and cannot be neglected, BlackMax results for rotating BHs should not be trusted.}. A
quick look at BH production at the LHC reveals the following. According to Thorne's hoop conjecture
\cite{hoop}, a BH of mass $M$ is formed when an object is compressed in all directions such that
\begin{equation*}
C < 2~\pi~R_{s}(M),
\label{thorne}
\end{equation*}
where C is the circumference of the region where the object is compacted into and $R_{s}$ is the
Schwarzschild radius for a BH of mass $M$. An upper limit on the BH mass is obtained by assuming no
gravitational energy loss at formation, corresponding to the black disk (BD) cross-section
$\sigma_{BD}=\pi R_{s}^2$. A more realistic model assumes that all the CM energy is not available
for BH formation, some being lost as gravitational radiation (see Ref.\ \cite{Cardoso:2005jq} for a
more detailed discussion). To estimate the energy loss, the colliding particles are treated as two
Aichelburg-Sexl shock waves \cite{Aichelburg:1970dh}; the overlap of the shock waves forms a
trapped-surface (TS) which sets a lower limit to the mass of the BH \cite{Yoshino:2002tx}. (For an
alternative estimate of the collisional gravitational loss, see Ref.\ \cite{Berti:2003si}.) The
cross-section at the LHC involves summing up the contributions from all the initial partons. The
cross section in the TS scenario is
\begin{eqnarray*}
\sigma_{pp \to BH}(s,n) =\sum_{ij}\int_0^1 2z dz
\int_{x_m}^1dx\,\int_x^1 \frac{dx'}{x'}\, f_i(x',Q)f_j(x/x',Q)\,F\,\sigma_{BD}(xs,n)\,,
\label{totcross}
\end{eqnarray*}
where $Q$ is four-momentum transfer squared, $f_i(x,Q)$ are the parton distribution functions, $z$
is the normalized impact parameter, and $F$ is a form factor. The cutoff in $x$ is related to the
minimum allowed mass of the object, $M_{min}$, and the fraction of center-of-mass energy trapped in
the BH, $y(z)$, by $x_m=M_{min}^2/[sy^2(z)]$. TeV BHs may carry electric or color charge and
angular momentum. Immediately after formation, they are expected to decay through loss of excess
multipole moments (balding phase), gravitational + Hawking radiation \cite{Hawking:1974sw}
(evaporation phase) and final $n$-body decay or remnant production (Planck phase). SM particles are
emitted on the brane and can be detected \cite{Emparan:2000rs}. Since the balding phase is poorly
understood, simulations neglect the energy loss in this phase. The description of the evaporation
phase is also approximated; since emissivities of rotating BHs are not known for all fields, BH
generators use greybody factors for non-rotating BHs \cite{Cardoso:2005mh,Kanti:2002nr}. In
CATFISH, the total decay multiplicity is \cite{Cavaglia:2006uk}
\begin{equation*}
N=\frac{(n+1)S}{4\pi}\,\frac{\sum_i c_i{\cal R}_i\Gamma_{{\cal R}_i}}
{\sum_j c_j{\cal P}_j\Gamma_{{\cal P}_j}}\,,
\label{multin}
\end{equation*}
where $c_i$ are the degrees of freedom of species $i$, $\Gamma_{{\cal P}_i}$ and $\Gamma_{{\cal
R}_i}$ are the relative emissivities of Ref.\ \cite{Cardoso:2005mh}, $S$ is the initial entropy of
the BH, and  ${\cal P}_s$ and ${\cal R}_s$ are spin-dependent normalization factors. A more
detailed discussion of the evaporation and Planck phases of TeV BHs can be found in
Refs.~\cite{Cavaglia:2002si}.

Dilepton production in BH events differ greatly from dilepton production in the MSSM. Unlike SUSY,
there is no single process of dilepton production; dileptons are either produced by the BH directly
or by the decay of heavier particles such as the $Z_0$ boson, $t\bar{t}$ pairs of a combination of
the two. Therefore, the BH dilepton invariant mass does not show a sharp cut-off at high energy.  
\section{Event Simulations\label{evt_sim}}
SUSY simulations are carried out using a combination of \texttt{ISAJET} and \texttt{PYTHIA}, with
the former generating the mass spectrum. BH simulations are carried out using the  \texttt{CATFISH}
Monte Carlo generator. The setup for each simulation is summarized below:
\begin{itemize}
\item SUSY:
\begin{itemize}
\item The MSSM mass spectrum is generated with \texttt{ISAJET}
(ver.\ 7.75);
\item The mass spectra in SLHA format is fed into \texttt{PYTHIA};
\item All SUSY processes except SM Higgs production are simulated;
\item Unstable SM particles and sparticles are hadronized or decayed with \texttt{PYTHIA}.
\end{itemize}
\item BHs:
\begin{itemize}
\item The cross section for a BH event is calculated in the center-of-mass frame;
\item The initial BH mass is sampled from the differential cross section;
\item The BH is decayed through Hawking mechanism and final $n$-body event (or remnant);
\item Unstable quanta are hadronized or decayed with \texttt{PYTHIA}.
\end{itemize} 
\end{itemize} 
The benchmark model for SUSY is LHC point A. The parameters for the BH benchmark model are
fundamental Planck scale $M_\star=1$ TeV, minimum BH mass $M_{min}=2$ TeV, classical-to-quantum
threshold $Q_{min}=1$ TeV, six extra-dimensions ($n=6$) and two-body final decay ($n_p=2$).
Particles produced in the initial-radiation phase are removed by imposing $P_T$ cuts of $5$ GeV and
$15$ GeV for leptons and photons+hadrons, respectively \cite{Cavaglia:2006uk}. 
\section{Event Analysis\label{evt_analysis}}
In this section, we first use event shape variables to discriminate SUSY and BHs models. We then
complement these results by looking at isolated dilepton events. The salient features of this
analysis is that BH events tend to be more spherical than SUSY events due to the spherical nature
of the Hawking radiation. This is specially evident for high-mass BHs. The formation of a stable BH
remnant at the end of the evaporation phase also helps to discriminate MSSM and BH events because
of the large amount of energy which is carried away by the remnant. Isolated dilepton events
provide a further powerful means to distinguish the two models. This is due to the fact that
leptons are rarely emitted by BHs (the hadron-to-lepton ratio is approximately 5:1) and are
uncorrelated; they can be emitted at any angle w.r.t.\ beam axis, whereas SUSY dileptons typically
originate from a single decay chain.
\subsection{\missPT\ and event shape variables\label{shape}}
\begin{figure*}[ht]
\centerline{\includegraphics*[width=0.50\textwidth]{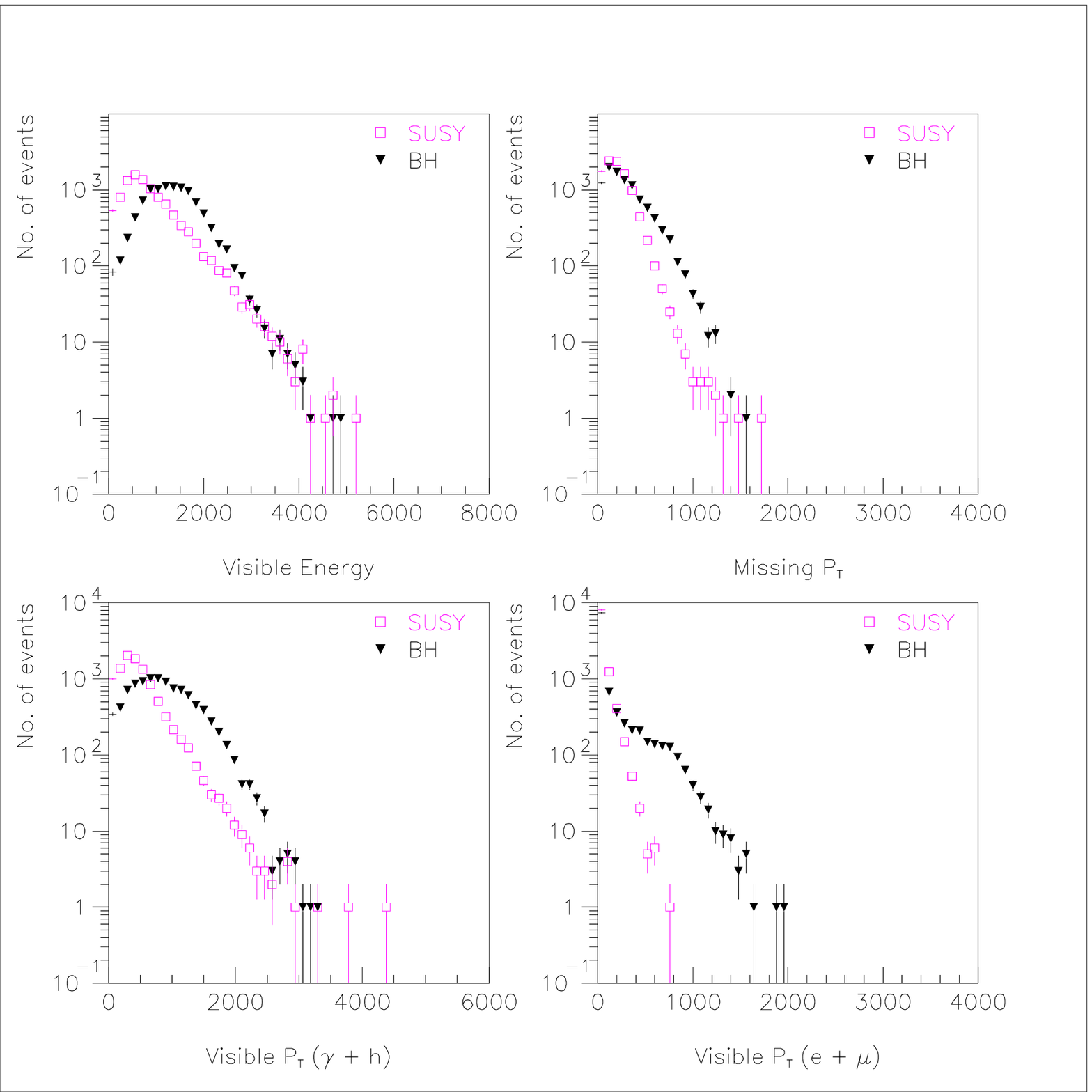}}
\caption{Comparison of 10,000 SUSY and BH benchmark events at the LHC. Visible energy and
missing transverse momentum \missPT\ (top panels) are comparable due to the presence of
invisible channels in both models. Leptons with large transverse momentum provide instead an
effective discriminator (bottom right panel).}
\label{FIG2}
\end{figure*}

Figure \ref{FIG2} shows visible energy, missing transverse momentum, and transverse momentum of
leptons and hadrons+photons for 10,000 MSSM and BH benchmark events. Even in the absence of a BH
remnant, the amount of visible energy and \missPT\ is comparable for the two scenarios. This is due
to the presence of invisible channels in both models: the LSP for SUSY and neutrinos+gravitons for
the BH. The flavor of the decay products is a better discriminator. MSSM interactions do not
produce leptons with energy above the TeV since isolated leptons are produced by the decay of
sparticles with typical energy of less than a few hundred GeV. On the contrary, quanta produced in
the BH decay are characterized by an average energy $E\sim M/N$, where the multiplicity $N$ is less
than 10 for typical BHs at the LHC. Since Hawking evaporation does not distinguish leptons from
hadrons, hard leptons with energy up to several TeV are likely to be produced during the BH decay.
This suggests that isolated leptons may provide a powerful means to discriminate the two models.
This is indeed the case, as we shall see in the next section.

\begin{figure*}[ht]
\centerline{\null\hfill
    \includegraphics*[width=0.5\textwidth]{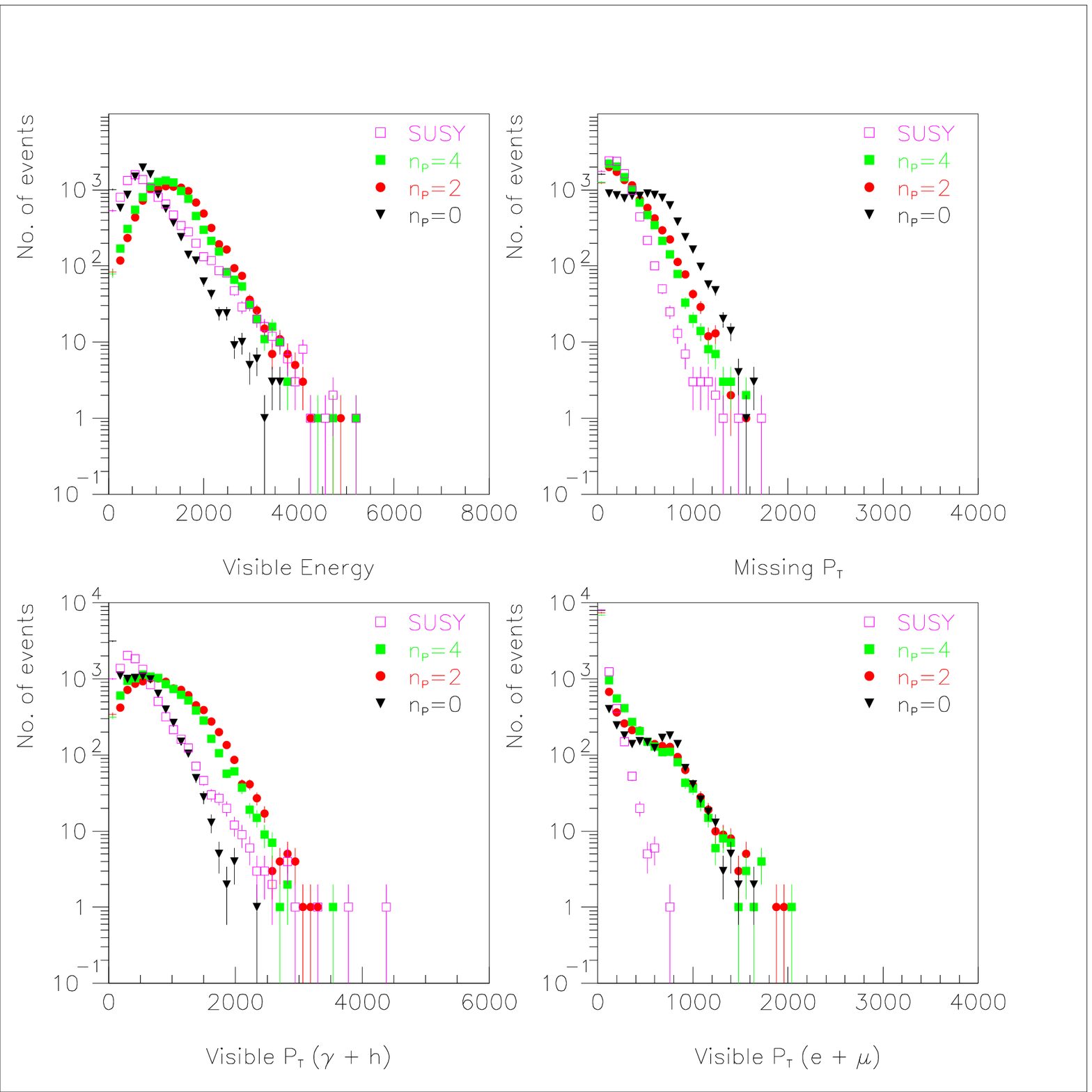}
    \hfill
    \includegraphics*[width=0.5\textwidth]{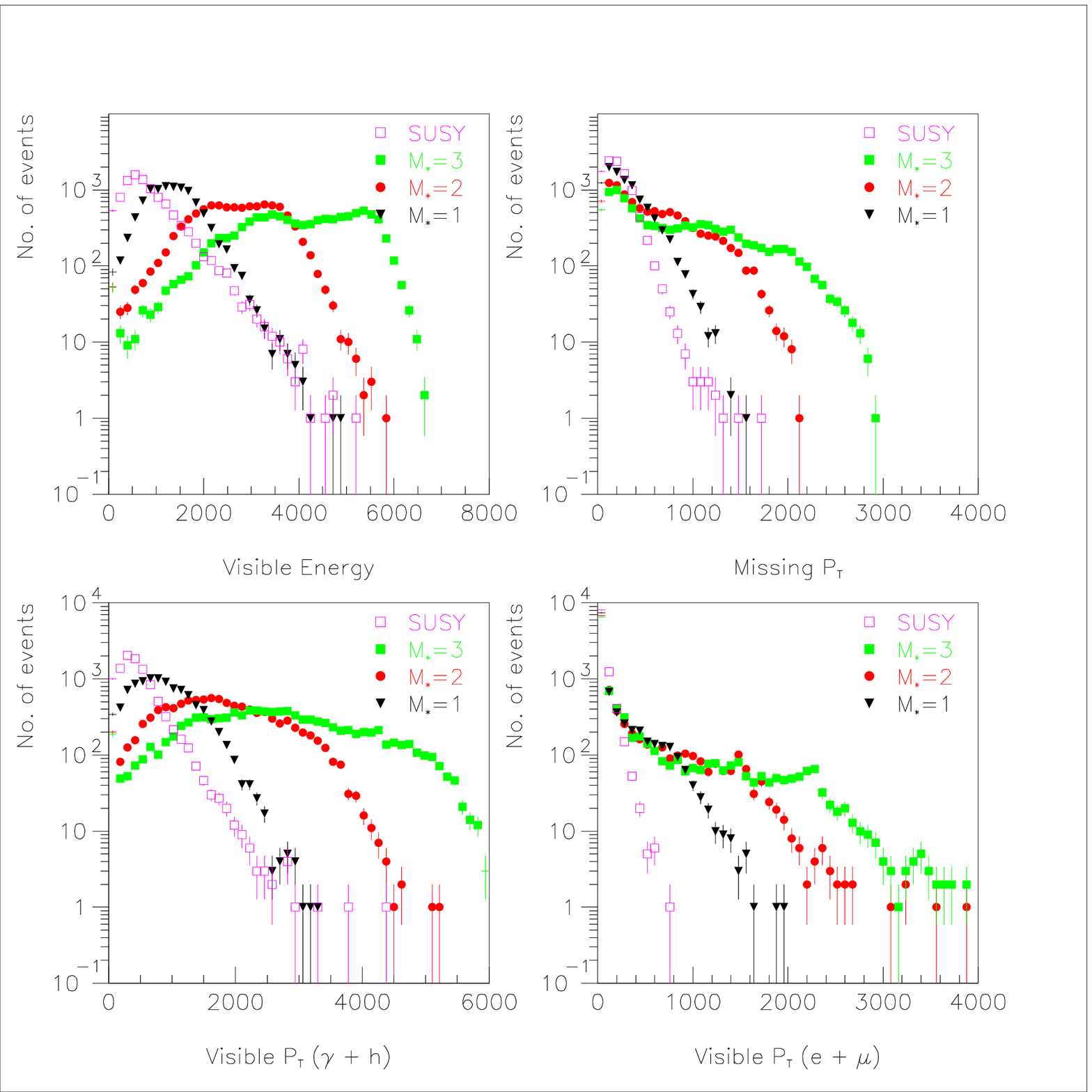}
    \null\hfill}
\caption{Distribution of visible energy, \missPT\ and transverse momenta of leptons and
hadrons+photons. SUSY plots are shown as pink open squares. The four plots in the left panel show
the effect of different decay modes in the Planck phase of ten-dimensional BHs: remnant formation
($n_p=0$, filled black triangles), two-body decay ($n_p=2$, filled red circles) and four-body decay
($n_p=4$, filled green squares). The fundamental Planck scale is $M_\star=1$ TeV. The four plots in
the right panel show the effect of varying the fundamental Planck scale: $M_\star=1$ TeV (filled
black triangles), $M_\star=2$ TeV (filled red circles) and $M_\star=3$ TeV (filled green squares).
The ten-dimensional BHs decay in two hard quanta at the end of the evaporation phase.}    
\label{FIG3}
\end{figure*}

The four plots in the left panel of Fig.\ \ref{FIG3} show how variations in the BH Planck phase
affect the observables of Fig.\ \ref{FIG2}. The plots compare two- and four-body decays to the
formation of a BH remnant ($n_p=0$). By the time the remnant has formed, the BH is expected to have
shed electric and color charges. (See, however, Ref.\ \cite{Koch:2005ks} for an alternative
scenario.) If this is the case, the BH remnant is undetectable and a source of missing transverse
momentum in addition to neutrinos and gravitons which are emitted during the Hawking evaporation
phase. This leads to a larger difference in \missPT\ between the MSSM and BH models. The visible
transverse momentum in hadrons+photons is sensibly reduced in the presence of a BH remnant; the
latter carries away energy which otherwise would have been emitted in visible channels (mostly
hadrons) during the BH decay phase. It is interesting to note that the amount of transverse
momentum in the leptonic channel is essentially unaffected by the presence of a BH remnant. This is
due to the fact that leptons are rarer than hadrons in the BH decay phase; variations in the energy
distribution of the leptonic channel are thus suppressed compared to the hadronic channel. Changes
in the number of final Planckian hard quanta do not produce significant differences in the
distributions; more quanta of lower energy behave statistically like less quanta with higher
energy. Provided that the BH decays at the end of the Hawking phase, it is thus safe to set the
number of Planckian quanta to $n_p=2$ or $n_p=4$, although BHs may decay in different numbers of
particles on a event-to-event basis. Variations in the classical-to-quantum threshold $Q_{min}$ are
also not expected to cause significant differences in the energy/momentum distributions. A higher
threshold increases the emission in the Planck phase while decreasing Hawking radiation. Since
these phases differs only in relative greybody factors, the effect is too small to be detected. 

The four plots in the right panel of Fig.\ \ref{FIG3} show visible energy, \missPT, and visible
transverse momenta of leptons and hadrons+photons for different values of the fundamental Planck
scale. Higher values of $M_\star$ lead to more massive BHs, i.e.\ higher multiplicity and more
energetic quanta. This causes a significant increase in missing and visible momenta. If the value
of the fundamental Planck scale happens to be large, BHs are likely to be found and easily
distinguished from SUSY through detection of highly-energetic isolated leptons and hadronic jets.
Missing transverse momentum of several TeV would also be observed.

\begin{figure*}[ht]
\centerline{\null\hfill
    \includegraphics*[width=0.5\textwidth]{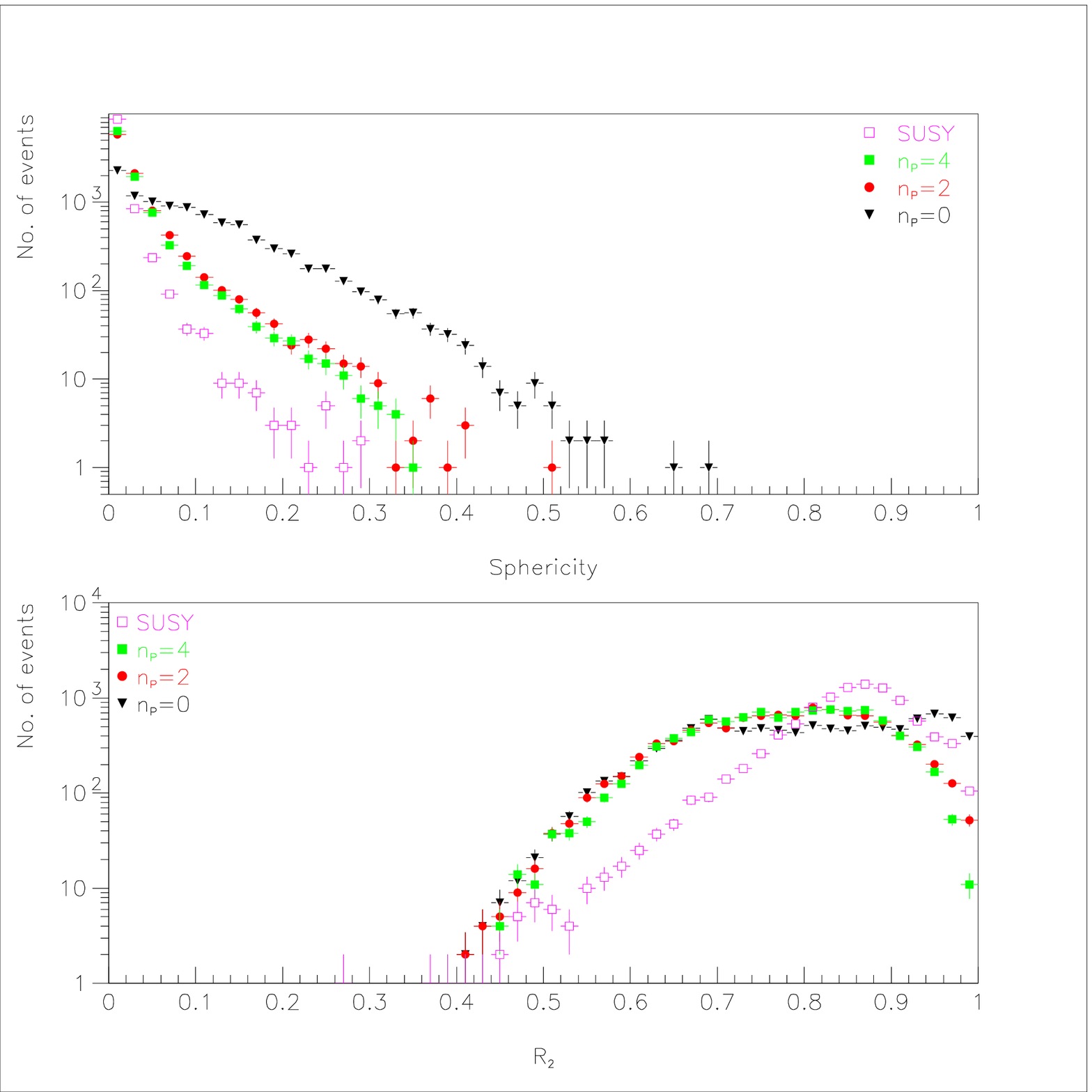}
    \hfill
    \includegraphics*[width=0.5\textwidth]{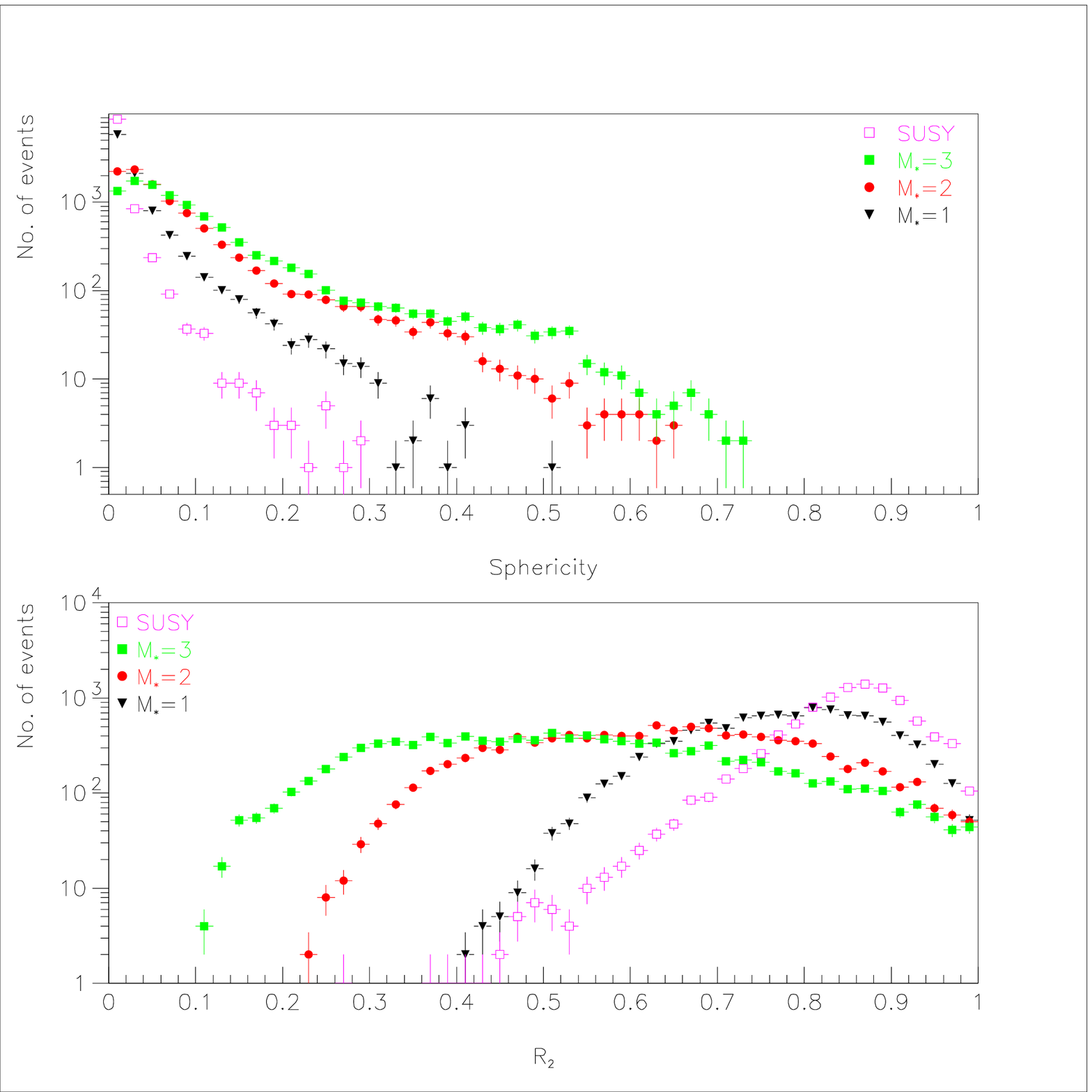}
    \null\hfill}
\caption{Sphericity (top panels) and $2^{\rm nd}$ Fox-Wolfram moment (bottom panels) for 10,000 BH
and MSSM events (pink open squares). The left panels show the effect of different Planckian decay
modes: BH remnant (filled black triangles), two-body decay (filled red circles) and four-body decay
(filled green squares). The fundamental scale is $M_\star=1$ TeV and the number of extra dimensions
is six. The right panels show the effect of different fundamental scales: $M_\star=1$ TeV (filled
black triangles), 2 TeV (filled red circles) and 3 TeV (filled green squares). The ten-dimensional
BHs decay in two quanta at the end of the Hawking phase.}  
\label{FIG4}
\end{figure*}

Event shape variables such as sphericity and $2^{\rm nd}$ Fox-Wolfram moment can be used to
complement the above analysis. BH events are more spherical because of the nature of Hawking
radiation and the ``jetty'' nature of SUSY decays. Formation of a BH remnant and high values of the
fundamental scale lead to significant higher sphericity than SUSY (top panels of Fig.\ \ref{FIG4}).
The $2^{\rm nd}$ Fox-Wolfram moment (bottom panels of Fig.\ \ref{FIG4}) is stable versus changes in
the BH Planck phase and provides a good MSSM/BH discriminator. BH models with higher $M_\star$ can
be differentiated more easily from the MSSM.

\begin{figure*}[ht]
\centerline{\null\hfill
    \includegraphics*[width=0.33\textwidth]{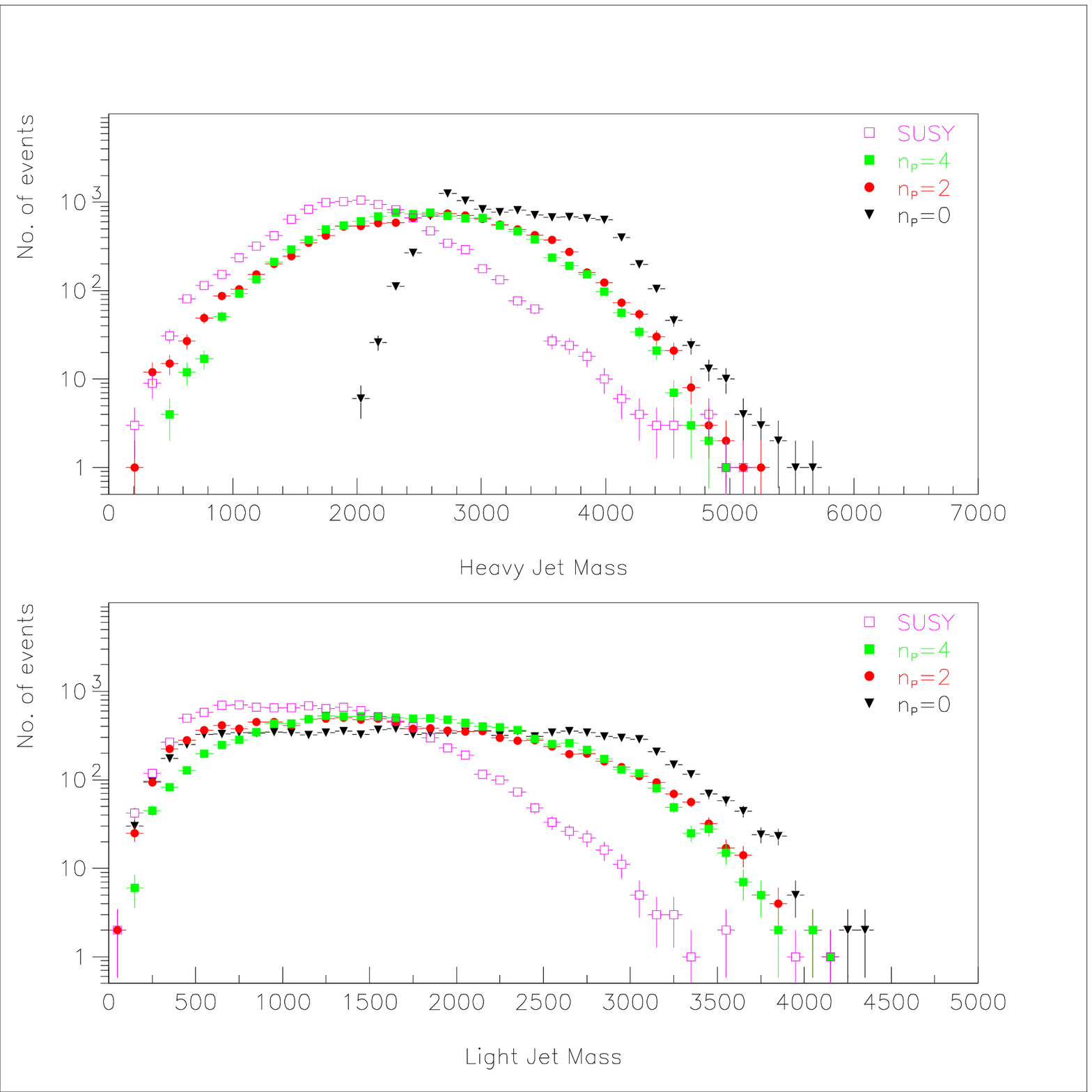}
    \hfill
    \includegraphics*[width=0.33\textwidth]{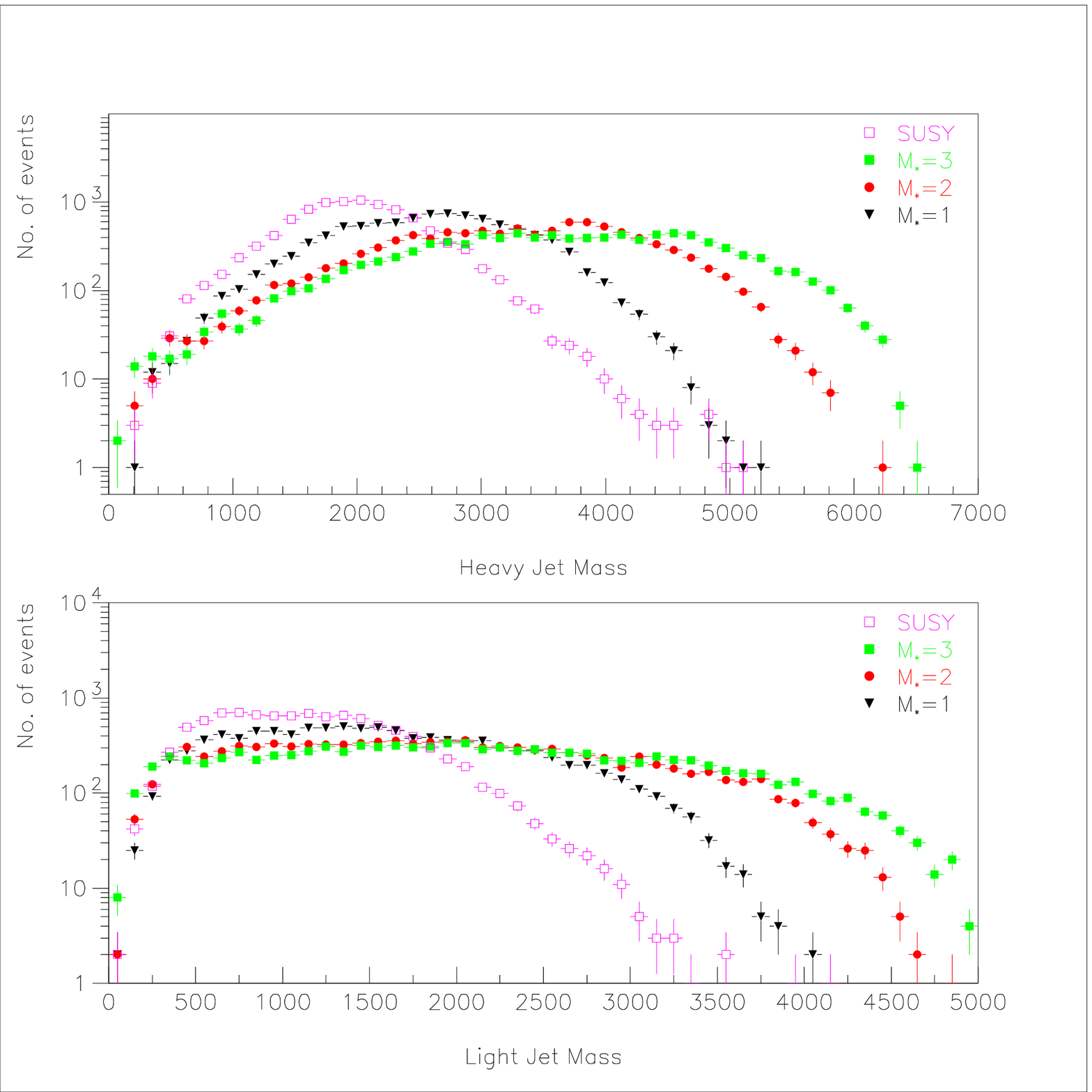}
    \hfill
    \includegraphics*[width=0.33\textwidth]{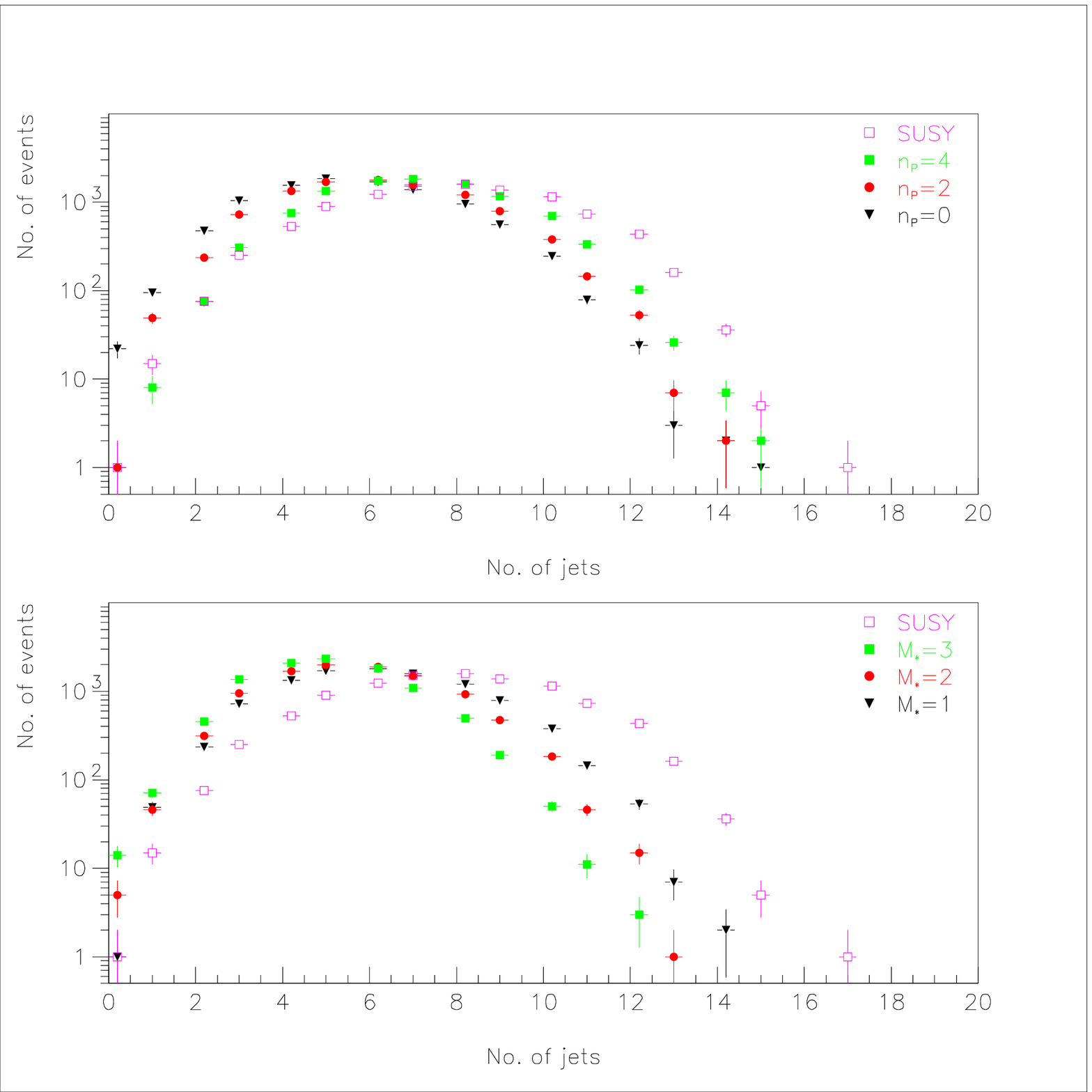}
    \null\hfill}
\caption{Light and heavy jet masses (four leftmost panels) and number of jets (right panels) for
10,000 BH and MSSM events. Symbols are like in previous figures.} 
\label{FIG5}
\end{figure*}

Similar conclusions can be reached by looking at jet masses and number of jets. The MSSM generates
more and lighter jets than the BH model due to copious production of quarks (Fig.\ \ref{FIG5}). The
difference is again specially significant for high values of $M_\star$ and in the presence of BH
remnants. Absence of sub-$Q_{min}$ hard jets could provide strong evidence for BH remnant
production. (See the suppression of heavy jets below the classical-to-quantum threshold $Q_{min}=2$
TeV in the top leftmost panel of Fig.~\ref{FIG5}.)

\subsection{Event analysis using high-$P_T$ dileptons \label{dilep}}
The use of isolated dileptons as SUSY signature has been extensively discussed in the literature
\cite{Hinchliffe:1996iu,Abdullin:1998nv}. Although their production is not as high as colored
particles, high-energy isolated leptons provide a cleaner environment by allowing the removal of
the QCD background. Moreover, since most of BHs produced at the LHC are expected to be very light,
multi-particle analysis may not provide the most effective  discriminators \cite{Cavaglia:2006uk,
Meade:2007sz}. The study of leptonic final states alleviates this problem.

The dominant MSSM interaction for opposite-sign, same-flavor (OSSF) dileptons at LHC point A is
\cite{atlas111}
\begin{eqnarray*}
\tilde{\chi}_{2}^{0} \rightarrow  l^{\pm} &\tilde{l} & \\
&\bentarrow&
l^{\mp}~ \tilde{\chi}_{1}^{0}
\label{chi2}
\end{eqnarray*}
with a branching ratio of 27\%. The maximum dilepton invariant mass for this interaction is
\begin{equation*}
M_{ll}^{max}=m_{\tilde{\chi}_{2}^0}
\left[\left(1-\frac{m_{\tilde{l}}^2}{m_{\tilde{\chi}_{2}^0}^2}\right)
\left(1-\frac{m_{\tilde{\chi}_{1}^0}^2}{m_{\tilde{l}}^2}\right)\right]^{1/2}
\sim 100~{\rm GeV}\,.
\label{mll}
\end{equation*}
The background for this process is due to SM decays of $W$, $Z$ bosons and top quarks. This
background can be removed by applying suitable cuts on transverse momentum and sphericity of the
leptons \cite{Bartl:1996dr}: 
\begin{itemize}
\item $P_{Tl} \ge 15$ GeV, $|\eta_l| < 2.5$;
\item Isolation cut, $\sum_l P_{Tl} < 7$ GeV in a cone of $R=0.2$,
\end{itemize}
where $P_{Tl}$ is the transverse momentum of the leptons, $R=\sqrt{\Delta\eta^2+\Delta\phi^2}$,
$\eta=-\ln[\tan(\theta/2)]$ is the pseudorapidity, and $\phi$ and $\theta$ are the azimuthal and
polar angles of the lepton w.r.t.\ beam axis, respectively.

Isolated leptons in BH decays come directly from the BH itself, from the decay of $Z_0$ bosons and
top quarks, or from a combination of the two. Since the branching ratio of $Z_0$ into leptons is
small, $\Gamma(l^{+}l^{-})/\Gamma_{\rm tot}\sim 0.034$ \cite{PDG}, and the decay of top quarks into
leptons is rare \cite{Abbott:1997fv}, production of OSSF dileptons is less frequent in the BH model
than in the MSSM. our analysis shows that an OSSF dilepton event occurs approximately every 100 BH
and 20 SUSY events, with a $\sim$ 1:5 ratio of BH-to-SUSY dilepton events at fixed luminosity.

\begin{figure*}[ht]
\centerline{\null\hfill
    \includegraphics*[width=0.5\textwidth]{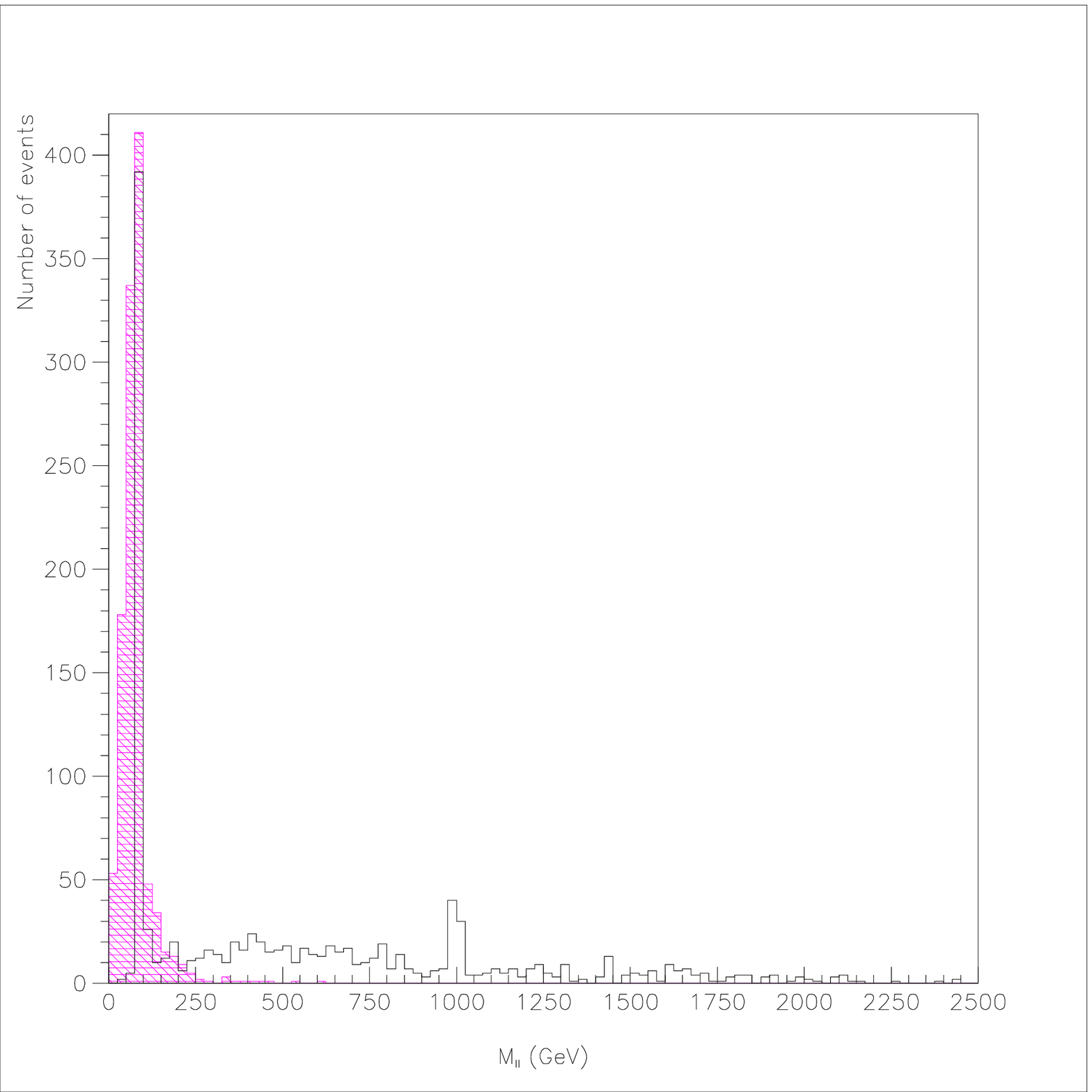}
    \hfill
    \includegraphics*[width=0.5\textwidth]{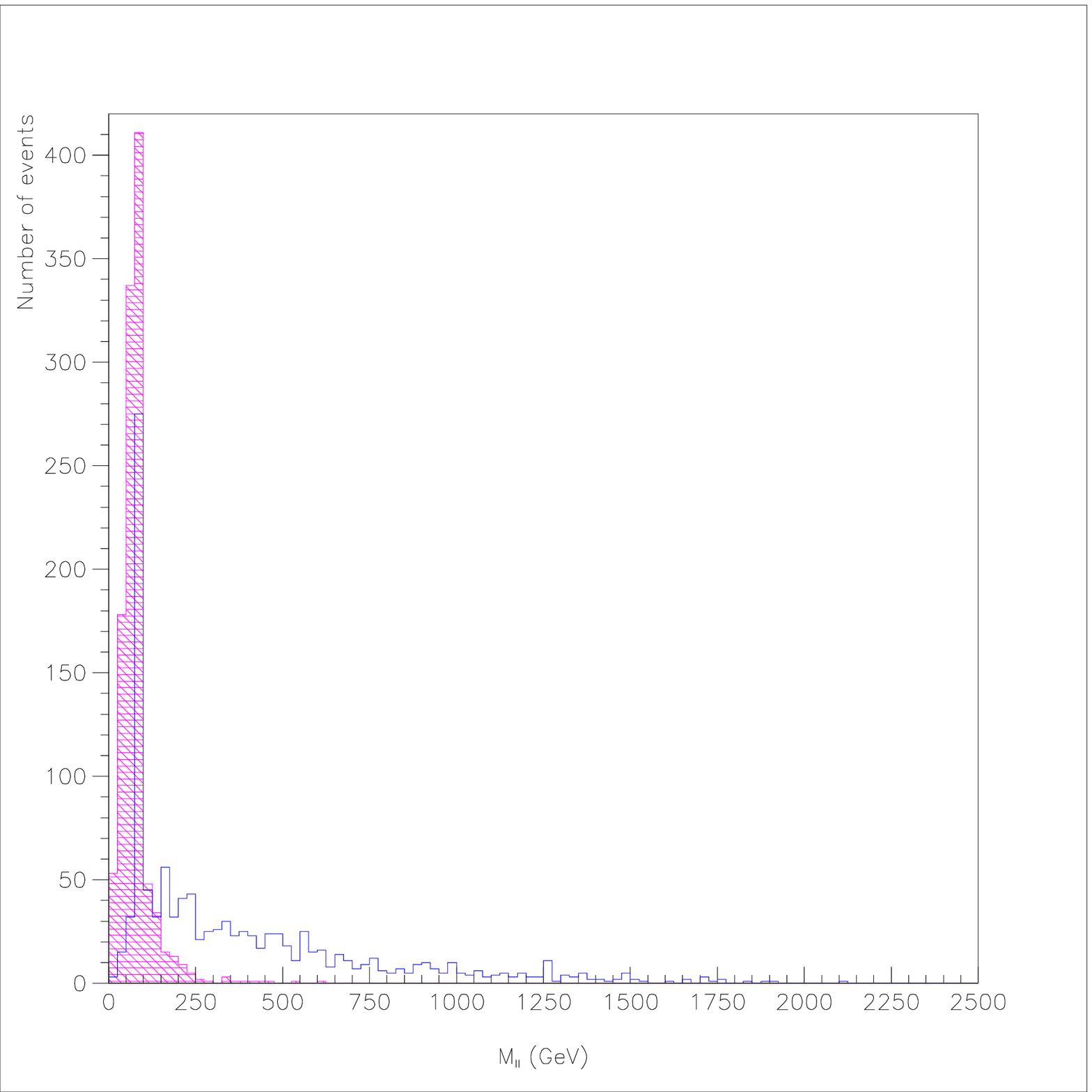}
    \null\hfill}
\caption{Invariant mass distribution (in GeV) for 1000 SUSY and BH OSSF dilepton events. The
SUSY distribution (shaded pink histogram) shows the typical endpoint due to the presence of the
LSP. The high-$P_T$ tail of the BH distribution is originated by uncorrelated lepton pairs
emitted during the Hawking evaporation phase. The final BH decay is in two-quanta (left panel)
or four-quanta (right panel).} 
\label{FIG6}
\end{figure*}

Figure \ref{FIG6} shows the dilepton invariant mass distribution for the MSSM (shaded pink plot)
and the BH model with final two- and four-body decay (left and right panels, respectively). As
expected, the SUSY distribution shows a sharp edge at $\sim$ 100 GeV \cite{Bartl:1996dr}. The BH
invariant mass distribution is characterized by a peak at $\sim 90$ GeV and a long tail up to
energy of several TeV. The peak is due to dileptons events produced from the decay of $Z_0$ bosons,
the dominant channel for OSSF dilepton production in BH. The tail is originated by uncorrelated
lepton pairs emitted directly by the BH or in top quark decays. The leptons are hard and the
reconstructed dilepton mass can have super-TeV values. The BH invariant mass for the two-body
Planckian decay shows a second, smaller peak at $\sim$ 1 TeV. This occurs because the BH at the end
of the Hawking phase may decay in two OSSF leptons, leading to a reconstructed dilepton invariant
mass equal to $Q_{\min}$. If the Planckian decay is a four-body process, the BH mass at the end of
the Hawking evaporation is distributed among four quanta. This produces a lower, smoother
reconstructed invariant mass.

\begin{figure*}[ht]
\centerline{\null\hfill
    \includegraphics*[width=0.5\textwidth]{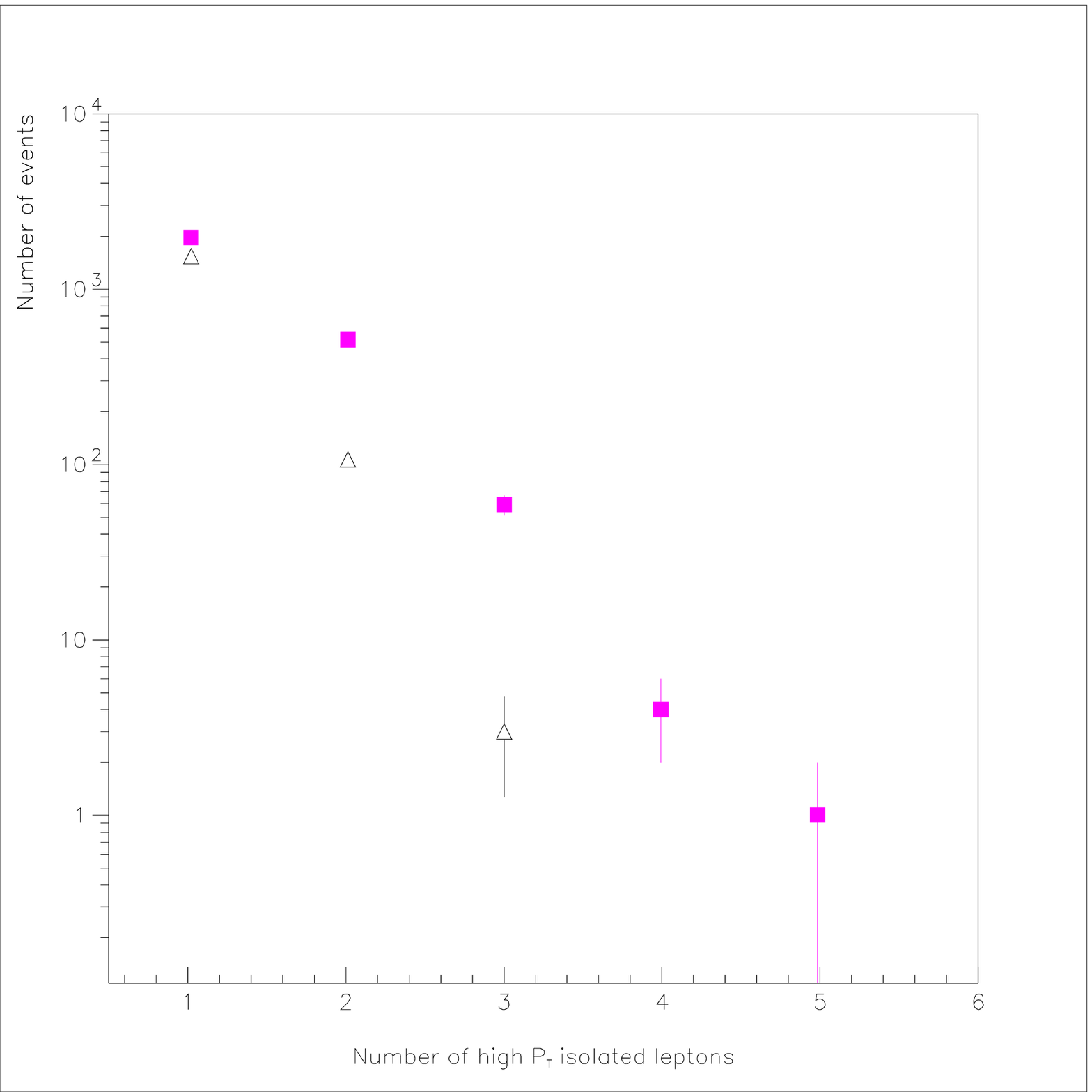}
    \hfill
    \includegraphics*[width=0.5\textwidth]{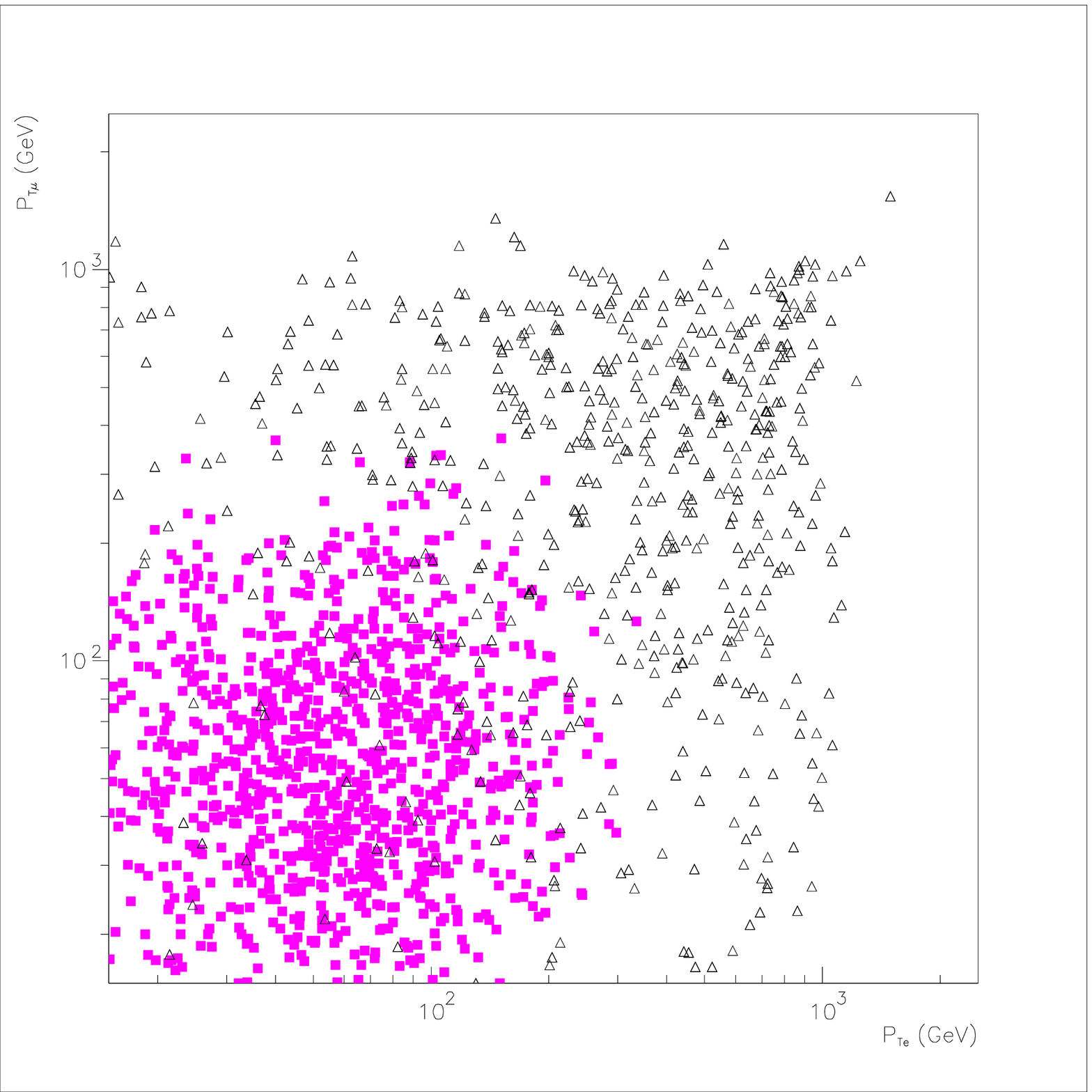}
    \hfill}
\caption{Left Panel: Histogram of the number of events with high-$P_T$ leptons for 10,000 MSSM
(pink filled squares) and BH interactions (black open triangles). The number of BH events with three
isolated leptons is smaller than the number of SUSY events by a factor of $\sim$ 20. The
probability of producing BH events with four or more leptons is virtually zero. Right Panel: $P_T$
scatter plot for $\sim$ 1000 isolated opposite-flavor dilepton events for SUSY (pink filled
squares) and BHs (black open triangles). BH leptons are harder than SUSY leptons and show a larger
spread in $P_T$.} 
\label{FIG7}
\end{figure*}

The number of isolated, high-$P_T$ leptons can also be used to complement the dilepton analysis
(left panel of  Fig.~\ref{FIG7}). SUSY events are capable of producing up to five isolated leptons
from the cascade decay of heavy sparticles. Events with $\tilde{\chi}_{2}^{0} \tilde{\chi}_{2}^{0}$
or $\tilde{\chi}_{1}^{\pm} \tilde{\chi}_{2}^{0}$ may produce four or three isolated leptons,
respectively \cite{Baer:1995tb}. On the contrary, events with three or more isolated leptons are
very suppressed in BH decays at the LHC energy. Although multilepton events are rare, there is very
little background and they could be effectively used to distinguish the MSSM and the BH model.
Other effective discriminators can be constructed by looking at dilepton events with same sign
and/or opposite-flavor leptons. The ``democratic'' nature of the BH decay makes all dilepton events
roughly equally probable, whereas the MSSM favors same-flavor dileptons. Presence of hard
opposite-flavor leptons is a clear indication of BH decay (right panel of Fig.~\ref{FIG7}). Our
analysis shows that 73\% of SUSY dilepton events are OSSF, compared to only 50\% in the BH model.
Conversely, opposite-flavor events are twice more frequent in the BH model (40\%) compared to the
MSSM (21\%). 

\section{Conclusions\label{concl}}
We have discussed and compared the signatures of the MSSM and the TeV-BH model at the LHC. A
thorough analysis of event-shape variables and dilepton events has shown that it is possible to
distinguish the two models. BH events are characterized by higher sphericity than SUSY processes.
If a BH remnant is formed at the end of the evaporation phase, missing $P_T$ and heavy jet mass are
effective signatures to discriminate BH formation from the MSSM. Although event-shape variables
alone cannot unequivocally discriminate between SUSY and BHs, their knowledge may prove useful when
combined with the analysis of the leptonic channel. Isolated dileptons could provide the ``smoking
gun'' for detecting BHs at the LHC. The BH dilepton invariant mass shows a tail at high energy
which is absent in the SM or MSSM. This analysis can be further strengthened by looking at the
number and flavor of isolated leptons.
\section*{Acknowledgments} 
The authors thank Lucien Cremaldi, Alakabha Datta, Romulus Godang, David Sanders and Don Summers
for fruitful discussions, Peter Skands for help with SLHA input file and Xerxes Tata for help on
\texttt{ISAJET}.  This work was supported (in part) by U.S.\ DoE contract DE-FG05-91ER40622.

\end{document}